\documentclass[journal=jacsat,manuscript=article]{achemso}

\usepackage[version=3]{mhchem} 
\usepackage[sort&compress,numbers,super]{natbib}
\usepackage{xcolor}
\usepackage{ulem}
\mciteErrorOnUnknownfalse

\author{Rachel Nixon}
\affiliation{Max Planck Institute for Chemical Physics of Solids, Dresden, Germany}
\altaffiliation{School of Chemistry, University of St Andrews, St Andrews, United Kingdom}

\author{Nazar Zaremba}
\affiliation{Max Planck Institute for Chemical Physics of Solids, Dresden, Germany}

\author{Samuel A. Adegboyega}
\affiliation{Department of Chemistry and Biochemistry, Florida State University, Tallahassee, United States}

\author{Andreas Leithe-Jasper}
\affiliation{Max Planck Institute for Chemical Physics of Solids, Dresden, Germany}

\author{Mitja Krnel}
\affiliation{Max Planck Institute for Chemical Physics of Solids, Dresden, Germany}

\author{Yurii Prots}
\affiliation{Max Planck Institute for Chemical Physics of Solids, Dresden, Germany}

\author{Ulrich Burkhardt}
\affiliation{Max Planck Institute for Chemical Physics of Solids, Dresden, Germany}

\author{J\"org Sichelschmidt}
\affiliation{Max Planck Institute for Chemical Physics of Solids, Dresden, Germany}

\author{Lucia Amidani}
\affiliation{ESRF, Grenoble, France}

\author{Fabio La Mattina}
\affiliation{Laboratory for Transport at Nanoscale Interfaces, Empa Swiss Federal Laboratories for Science and Technology, Dübendorf, Switzerland} 

\author{Michael Shatruk}
\affiliation{Department of Chemistry and Biochemistry, Florida State University, Tallahassee, United States}

\author{Alexander Shengelaya}
\affiliation{Department of Physics, Ivane Javakhishvili Tbilisi State University, Tbilisi, Georgia}
\altaffiliation{Andronikashvili Institute of Physics, Ivane Javakhishvili Tbilisi State University, Tbilisi, Georgia}

\author{Manuel Brando}
\affiliation{Max Planck Institute for Chemical Physics of Solids, Dresden, Germany}

\author{Eteri Svanidze}
\affiliation{Max Planck Institute for Chemical Physics of Solids, Dresden, Germany}
\email{svanidze@cpfs.mpg.de}


\title{Unusual magnetic order in Eu$_{10}$Hg$_{55}$}

\begin{document}

\begin{abstract}
In solid-state compounds, the valence of europium can sometimes be mixed -- which is especially favored in structures with several positions for the europium atoms. In this work, we study the Eu-based intermetallic noncentrosymmetric system Eu$_{10}$Hg$_{55}$ which has 65 atoms per unit cell and 4 distinct crystallographic positions for europium and 17 positions for mercury. Our detailed analysis of magnetism of large single crystals suggests that europium in Eu$_{10}$Hg$_{55}$ might be present in two valence states, resulting in a fragile magnetic ground state. Due to the cage-like structure with a large distance between the Eu atoms, those atoms are weakly ferromagnetically coupled and Eu$_{10}$Hg$_{55}$ orders at low temperatures, below $T_{1} = 5.5$ K, with a subsequent spin re-orientation at $T_{2} = 4.3$ K. There is no sign of magnetic frustration. Interestingly, the magnetic ordering of europium sub-lattices results in a magnetization pole reversal with a weak ferrimagnetic ground state. Additional magnetic phases can be induced by application of a modest external magnetic field.

\end{abstract}

\section*{Introduction}

Intermediate valence and mixed valence compounds offer a great playground for correlated electron physics and unusual bonding phenomena. The ions in rare-earth-based materials are typically in the trivalent state. However, in some compounds cerium may have an oxidation state of 3$^+$ or 4$^+$ while samarium, europium, thulium, and ytterbium may have oxidation state of 2$^+$ or 3$^+$. In fact even among rare-earths, europium is the odd one out -- it mostly exhibits 2$^+$ rather than 3$^+$ oxidation state favored by the other rare-earths \cite{Engel2023, Lawrence1981}. Furthermore, europium is significantly less abundant and has lower melting temperature and density compared to its neighbors in the Periodic Table. The similar size and stability of Eu$^{2+}$ explain the tendency of europium to substitute for Ca$^{2+}$ in minerals \cite{Weill1973, Bau1991}, serving as a pertinent geochemical marker. The flexible oxidation state of europium is also the reason behind many peculiar low-temperature properties of quantum materials containing europium -- in this sense, europium is similar to cerium, samarium, thulium, and ytterbium. While the sizes of the Eu$^{2+}$ and Eu$^{3+}$ ions differ (1.25 \AA~ vs. 1.066 \AA~ for coordination of 8) \cite{Shannon1976}, the fast electronic fluctuations frequently make the coexistence of both species possible. Additionally, some structures can promote sites with different volumes. Moreover, the mixed valence can either be static or dynamic, with the latter resulting in abrupt changes in structure and electronic properties. This has prompted many mixed valence compounds containing europium to be investigated over the past decades \cite{Honda2017, Varma1976, Vitins1977, Jakubcova2009, Zeuner2009, Felser1997, Lossau1989, Radzieowski2018, Wickleder2002, Felser1997new, Furuuchi2004, Jakubcova2007, Podgorska2025, Walicka2025, Artmann1996, Gumeniuk2016, Onuki2017, Nowik1983}.

While mercury is frequently associated with superconductivity  \cite{Onnes1911, Pelloquin1993, Carlson2004, Bussmann2007, Bryntse1997, Tresca2022}, its large spin-orbit coupling has been shown to promote emergence of peculiar topological states \cite{Konig2007, Virot2013, Dumett2020, Virot2011, Forcella2025, Mutter2004}. Associated experimental difficulties such as toxicity and air-sensitivity \cite{Hoch2019, Beletskaya2019} can be solved by utilizing a confined laboratory and adapted measurement methods \cite{LeitheJasper2006, Svanidze2019, Prots2022La, Prots2022U, Witthaut2023, Nixon2025, Nixon2024}. Recent synergetic efforts between chemical \cite{Hoch2008, Nusser2022, Tambornino2015, Tambornino2015_badmetal, Tambornino2015_Gd14Hg51, Wendorff2018, Wendorff2012, Wendorff2015, Falk2020, Schwarz2017, Tambornino2016, Hoch2012, Hoch2019, Bebout2006, Tkachuk2008} and physical \cite{Svanidze2019, Prots2022La, Prots2022U, Witthaut2023, Nixon2025, Nixon2024} investigations revealed that many new materials can be discovered -- spanning whole range of chemical complexity and ground states, ranging from magnets to superconductors.

In this work, we present magnetic properties of the noncentrosymmetric Eu$_{10}$Hg$_{55}$ compound, whose complex crystal structure lies at the origin of the intricate magnetic phase diagram. The average valence of Eu is ranging from 2 to 2.2, placing Eu$_{10}$Hg$_{55}$ in a family of inhomogeneous mixed-valence systems~\cite{Lawrence1981}. Magnetic order occurs below 5.5 K. The coordination environment of Eu in Eu$_{10}$Hg$_{55}$ is large, with coordination number (CN) of 14-16. This links Eu$_{10}$Hg$_{55}$ to magnetically ordered cluster-like compounds such as EuCd$_{11}$ and UCd$_{11}$ (CN = 20, AFM below $T_N$ = 2.7 K \cite{Buschow1976, Nakamura2014} and 5 K \cite{Fisk1984, Zaremba2022, Cafasso1963, Misiuk1973}, respectively), U$_{23}$Hg$_{88}$ (CN = 14-16, AFM below $T_N = 2.2$ K \cite{Svanidze2019}), and U$_2$Zn$_{17}$ (CN = 19, AFM below $T_N = 9.7$ K \cite{Misiuk1973}).

\section*{Analysis of crystal structure and magnetic properties}

\begin{figure}[t!]
\includegraphics[width=\columnwidth]{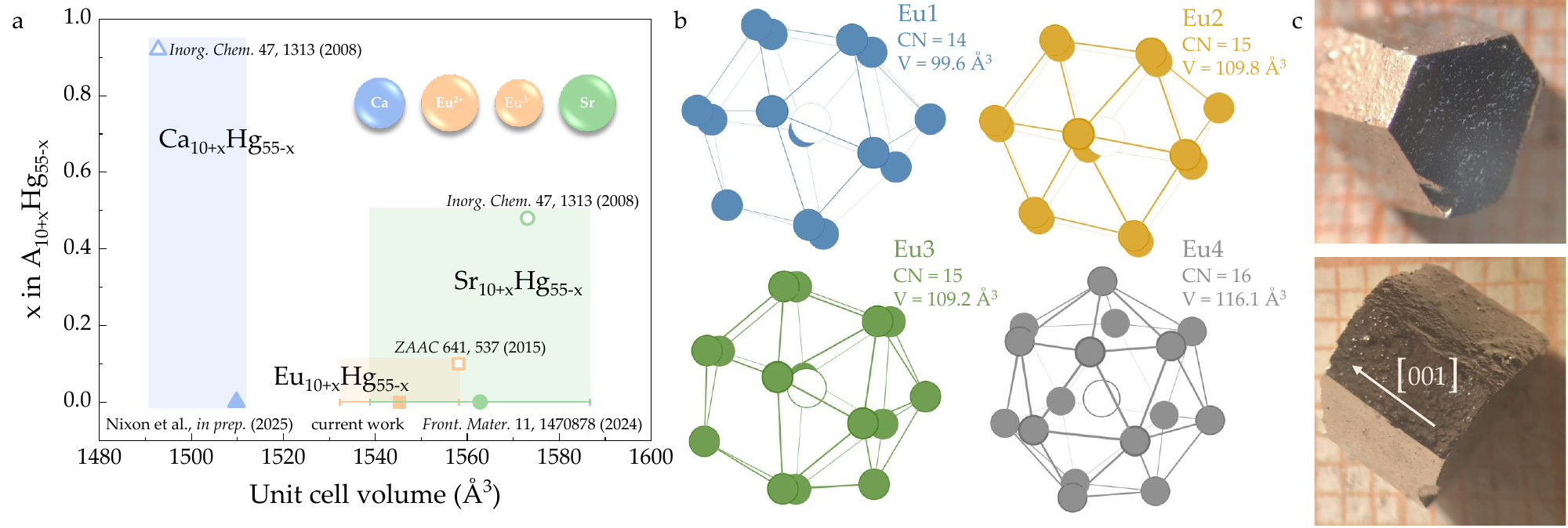}
\caption{(a) Small variations in the stoichiometry of the A$_{10}$Hg$_{55}$ family (A = Ca, Sr or Eu) are driven by mixed occupancy of one of the A1 position \cite{Tkachuk2008, Tambornino2015, Nixon2024}. For superconducting A = Ca and Sr (blue and green), these defects do not change the value of $T_c$ much -- on the order of 10\%  for $x \leq 0.5$ \cite{Nixon2024, Nixon2025}. For the case of Eu$_{10}$Hg$_{55}$, homogeneity range is significantly smaller (orange region), compared to the alkali elements. (b) The complex magnetic phase diagram of Eu$_{10}$Hg$_{55}$ is driven by the four distinct crystallographic sites on which Eu atoms are located. The coordination of Eu varies between 14 (Eu1), 15 (Eu2 and Eu3), and 16 (Eu4). (c) Single crystals of Eu$_{10}$Hg$_{55}$, placed on mm-paper.}
\label{fig:Structure}
\end{figure}

Among europium-based materials that have been reported to crystallize in the noncentrosymmetric $P\overline{6}$ structure, Eu$_2$(Ni/Co)$_{12}$P$_7$ \cite{Jeitschko1980, Reehuis1989} (Zr$_2$Fe$_{12}$P$_7$ structure type), Eu$_7$F$_{12}$Cl$_2$ \cite{Reckeweg2014} and EuNaF$_4$ \cite{Zakaria1997} (NaNdF$_6$ structure type) display trivalent europium. Divalent europium has been reported for the Eu$_2$Mg$_3$Cu$_9$(As/P)$_7$ \cite{Zhu2016} and Eu$_2$Yb$_{1.11}$Mg$_{10.89}$Si$_7$ \cite{Vasquez2020} compounds (both Zr$_2$Fe$_{12}$P$_7$ structure type) as well as Eu$_5$In$_9$Pt$_7$ \cite{Heying2019}, EuBa$_6$Cl$_2$F$_{12}$ \cite{Kubel2000} and Eu$_7$Cl$_2$F$_{12}$ \cite{Reckeweg2014} (both Ba$_7$Cl$_2$F$_{12}$ structure type). The formation of the mercurides with the 10:55 stoichiometry has so far only been reported for Na \cite{Hoch2012, Hornfeck2015}, Ca \cite{Tkachuk2008}, Sr \cite{Tkachuk2008, Nixon2024}, Eu \cite{Tambornino2015}, and Yb \cite{Tambornino2017} -- see Fig.~\ref{fig:Structure}(a). Given similarity of Eu (1.066 and 1.25 \AA) to Ca (1.12 \AA) and Sr (1.26 \AA), it is likely that the homogeneity ranges for the three isostructural systems would be comparable.\footnote{This comparison does not include the Na and Yb analogues, since for those only lattice parameters refined from the single crystal diffraction experiments have been reported \cite{Hoch2012, Hornfeck2015, Tambornino2017}. Our preliminary work on these systems indicate a slightly different value for the volumes, which are still under investigation \cite{Nixon2025}.} 

\begin{figure*}
\includegraphics[width=\columnwidth]{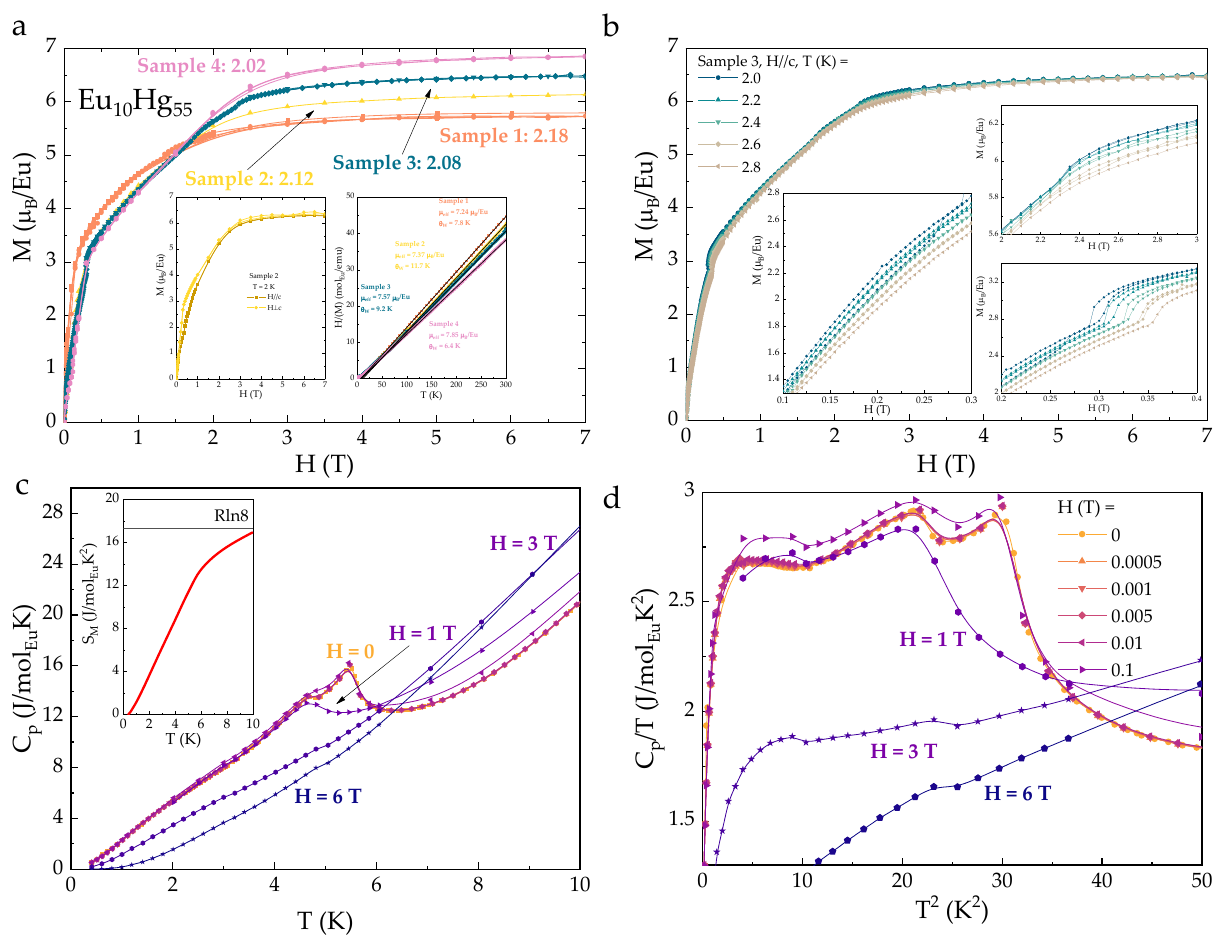}
\caption{Magnetic properties of Eu$_{10}$Hg$_{55}$. (a) Isotherms, taken at $T = 2 K$ saturate with a moment less than that expected for a purely Eu$^{2+}$ material ($\mu_{sat, theory}$ = 7 $\mu_B$). The value of saturated moment appears to be isotropic (see inset). Based on the value at $H = 7$ T, it is possible to estimate the relative ratio of Eu$^{2+}$ to Eu$^{3+}$, producing the overall mean value of the valence, which is shown for each of the samples. In the paramagnetic state, the value of the Weiss temperature and effective magnetic moment also varies between the samples (inset). (b) Temperature-dependent isotherms show a number of features (insets), which are used to construct the $H = T$ phase diagram for Eu$_{10}$Hg$_{55}$. (c) Entrance into magnetically ordered state below $T = 5.5$ K is marked by a sharp anomaly, which is gradually suppressed with magnetic field. The inset shows the entropy of Eu$_{10}$Hg$_{55}$, reaching R$ln$8 at $T = 10$ K. (d) The features, corresponding to transitions between different magnetic configurations are easier to track from the $C_p/T$ vs. $T^2$ data.}
\label{fig:MT}
\end{figure*}

Some deviations from the 10:55 stoichiometry are driven by mixed occupancy of the A4 position, with Eu analogue showing the smallest homogeneity range, see Fig.~\ref{fig:Structure}(a). Given the fact that the size of Eu$^{3+}$ (1.066 \AA~ \cite{Shannon1976}) is notably smaller than Eu$^{2+}$ (1.25 \AA~ \cite{Shannon1976}), it is likely that Eu$^{3+}$ occupies the Eu1 position (multiplicity of 3) -- see Fig.~\ref{fig:Structure} (middle). Since the other 7 Eu atoms are distributed among the Eu2 (3), Eu3 (3) and Eu4 (2)\footnote{According to previous single crystal refinement,\cite{Tambornino2015} this position is half-occupied by mercury, reducing the number of Eu atoms on this site to just 1.} positions, this suggests that the ratio of Eu$^{3+}$ to Eu$^{2+}$ in Eu$_{10}$Hg$_{55}$ is roughly 3:7. As discussed below, this, in principle, is in a good agreement with a mean value of the valence of 2.2 (Eu$^{3+}$ to Eu$^{2+}$ ratio of 2:8) that is estimated for Eu$_{10}$Hg$_{55}$ based on the $M(H)$ data (see Fig.~\ref{fig:MT})(a). Given high air-sensitivity, low melting temperature, and strong chemical reactivity of crystallites with freshly prepared surface, it was not possible to perform an in-depth analysis of crystal structure of Eu$_{10}$Hg$_{55}$, which would give a definitive Eu to Hg as well as Eu$^{2+}$ to Eu$^{3+}$ ratios.

\begin{figure}
\includegraphics[width=\columnwidth]{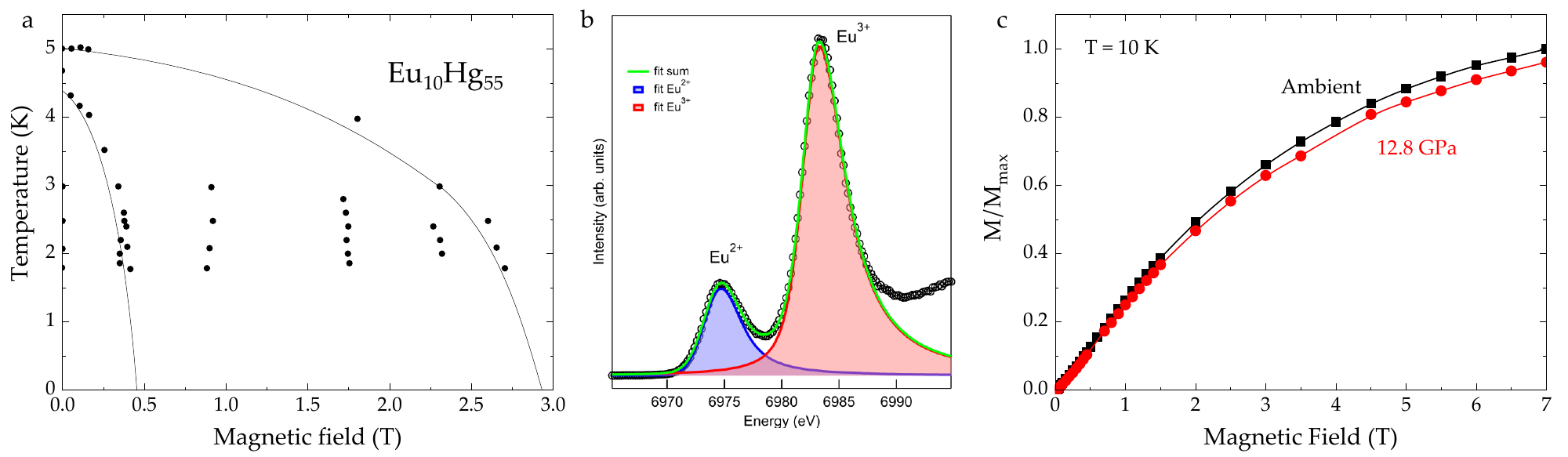}
\caption{(a) The $H-T$ phase diagram of Eu$_{10}$Hg$_{55}$ is likely driven by different magnetic sub-lattices of Eu. Within this phase diagram, the definitive assignment of magnetic configurations is not yet possible. (b) The XANES analysis indicates that both Eu$^{2+}$ and Eu$^{3+}$ are present in Eu$_{10}$Hg$_{55}$. However, the relative ratios are not quantitative, given that Eu$_{10}$Hg$_{55}$ decomposes into Hg and Eu$_2$O$_3$ (in which Eu is in the 3+ state) during the measurement. (c) Evolution of magnetic moment under ambient (black) and 12.8 GPa (red) pressure.}
\label{fig:PhaseD}
\end{figure}

The saturated magnetic moment expected for a purely Eu$^{2+}$ compound should amount to 7 $\mu_B$. As can be seen from Fig.~\ref{fig:MT}(a), all Eu$_{10}$Hg$_{55}$ samples show saturation of the $M(H)$ isotherm, albeit with a smaller value of the moment.\footnote{It is important to note that mixing of Eu and Hg on the Eu4 position can, in principle, produce compounds with more or less Eu, compared to the 10:55 ratio \cite{Tambornino2015}. However, to achieve a full 7 $\mu_B$ moment, a stoichiometry of Eu$_8$Hg$_{57}$ is needed (for Sample 1, for example), which would mean that the homogeneity range of this phase is far beyond that reported for other R/A$_{10-x}$Hg$_{55+x}$ phases -- see Fig.~\ref{fig:Structure}(a). For the rest of the analysis, we therefore use the Eu$_{10}$Hg$_{55}$ formula, since the magnetic behavior of this phase is driven by the ratio of the Eu$^{3+}$ to Eu$^{2+}$ species rather than by the ratio of Eu to Hg.} Based on the value of $\mu (H = 7~\text{T}, T = 2~\text{K})$, it is possible to estimate the ratio of Eu$^{3+}$ to Eu$^{2+}$ for each of the Eu$_{10}$Hg$_{55}$ samples, as summarized in Fig.~\ref{fig:MT}(a). The average Eu valence, listed for each of the samples, ranges from 2.02 (pink) to 2.18 (orange). It is important to note that while the low-field data shows some anisotropy (dark vs. light yellow of the inset), the saturated magnetic moment is isotropic. All samples exhibit Curie-Weiss behavior, as summarized in the inset. By fitting the inverse susceptibility above $T = 100$ K, a positive Weiss temperature $\theta_{W} = 6.4 - 11.7$ K is extracted. The $\theta_{W}$ temperature is positive, indicating ferromagnetic exchange interaction between the Eu atoms. In fact, the ordering temperature of Eu$_{10}$Hg$_{55}$ is about 5.5 K, but the ordered state is likely not purely ferromagnetic. This is not a surprise because the Eu atoms are in the 2+ state, with $L = 0$ and $S = 7/2$, which means isotropic moments with no effect of the crystalline electrical field because of the missing orbital moment. This implies that a secondary weak antiferromagnetic interaction can still drive the system to be antiferromagnetic. This is reminiscent of the EuCd$_{2}$P$_{2}$ system \cite{Nasrallah2024, Artmann1996, Sunko2022, Pakhira2022, Homes2023}, in which the relevant energy scale is ferromagnetic (positive $\theta_{W}$) and therefore the Eu moments align ferromagnetically within the plane of the tetragonal structure, but a very small antiferromagnetic coupling between the planes drives the system to an overall antiferromagnetic ordering. The value of the effective fluctuating moment $\mu_{eff}$ = 7.24-7.85 $\mu_{\textrm{B}}$, extracted from the Curie-Weiss fits, is close to the calculated value for Eu$^{2+}$ of 7.94 $\mu_{\textrm{B}}$. As summarized in Fig.~\ref{fig:MT}(b), magnetic isotherms show a number of features, which can be used to construct a preliminary phase diagram, see Fig.~\ref{fig:PhaseD}(a) below. We have included all features, even though they might be not a signature of phase transitions. Of course, a more detailed understanding of the magnetic structure in Eu$_{10}$Hg$_{55}$ would be desirable, however, neutron scattering experiments -- due to the large neutron cross sections of constituent elements -- are ruled out. It is, however, of interest to study this compound by $^{151}$Eu-M\"ossbauer spectroscopy and Eu L-edge X-ray magnetic circular dichroism spectroscopy, which should shed more light on this issue \cite{Furuuchi2004, Nowik1983, Jakubcova2007, Mullmann1997}. 

The entrance into magnetically ordered state below 5.5 K is also supported by the specific heat data of Eu$_{10}$Hg$_{55}$, shown in Fig.~\ref{fig:MT}(c). As seen in the inset, the full $R$ln8 is only recovered at $T = 10$ K, meaning that magnetic fluctuations persist above the ordering temperature. This makes the determination of the Sommerfeld coefficient $\gamma$ not possible for Eu$_{10}$Hg$_{55}$. The structural similarity of Eu$_{10}$Hg$_{55}$ to U$_{23}$Hg$_{88}$ \cite{Svanidze2019} suggests that the possibility of effective mass enhancement in former system should be investigated in more detail in the future.
\begin{figure}
\includegraphics[width=\columnwidth]{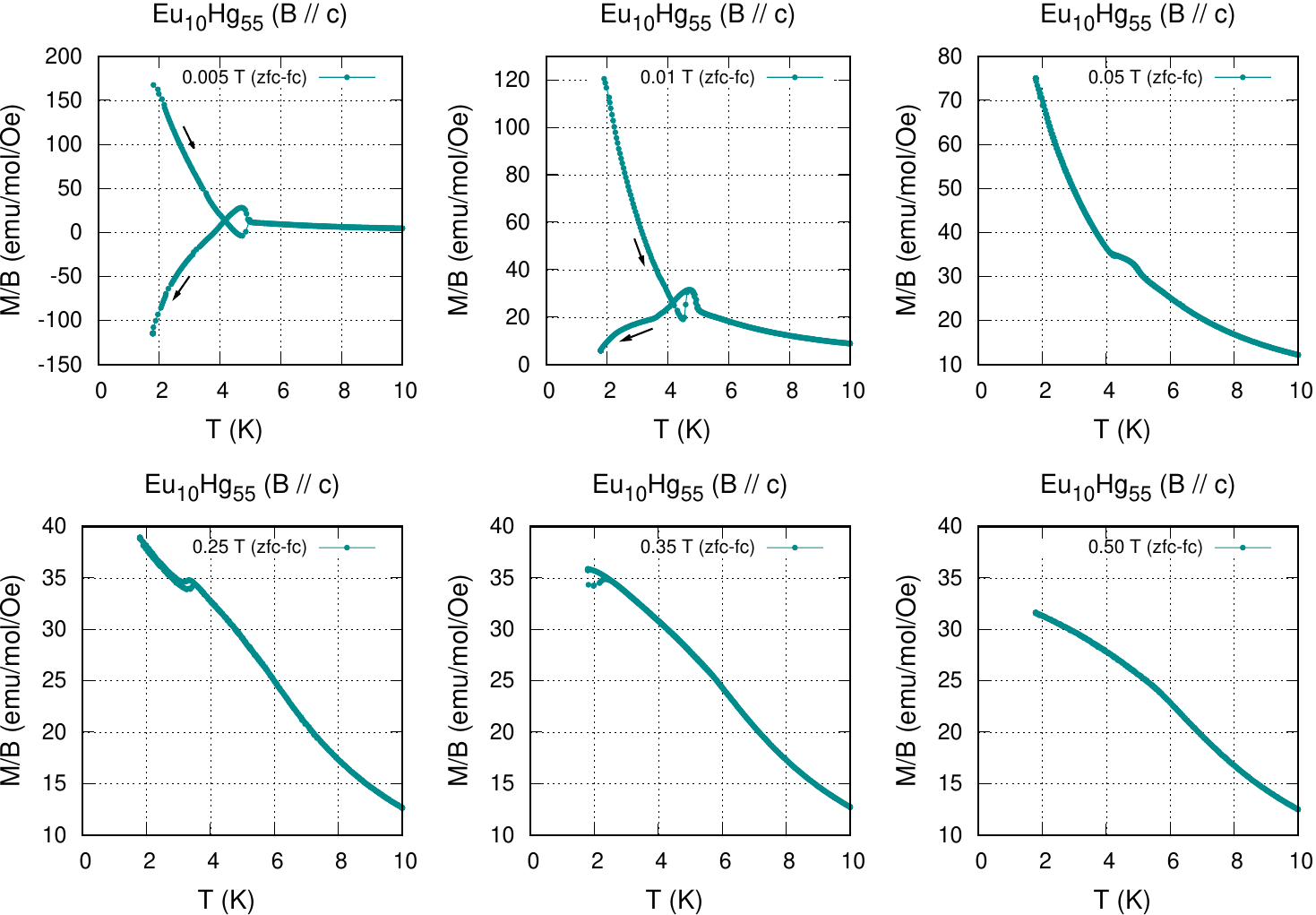}
\caption{Zero-field-cooled (zfc) field-cooled (fc) measurements of the magnetization of Sample 3 for six selected fields. The peculiar behavior observed at the lowest field of 0.005 T and with increasing fields is rare and is observed in pole reversal ferrimagnetic systems like Ni(HCOO)$_{2} \cdot$ H$_{2}$O\cite{kageyama2003}, Mn$_{3-x}$Ni$_{x}$BO$_{5}$\cite{bezmaternykh2014} or Pr$_{3}$Fe$_{3}$Sb$_{7}$~\cite{Pabst2023}.}
\label{fig:zfc-fc}
\end{figure}

The ordered state of Eu$_{10}$Hg$_{55}$ is unusual and fragile. This is demonstrated by a series of zero-field-cooled (zfc) and field-cooled (fc) measurements of the magnetization, taken at very small magnetic fields. Selected data are shown in Fig.~\ref{fig:zfc-fc}. Because of the very small fields used and the presence of remnant field in the superconducting magnet (a few mT), the absolute values of the are not precise. In the fc data with $B = 0.005$\,T, a very peculiar behavior is observed: First magnetization shows a small increase at $T_{1} \approx 4.98$ K and then a strong decrease below $T_{2} = 4.71$ K to negative values. In the zfc data, the behavior is opposite and very symmetric. This is reminiscent of the behavior of the magnetization pole reversal of ferrimagnetic systems like Ni(HCOO)$_{2} \cdot$ H$_{2}$O\cite{kageyama2003}, Mn$_{3-x}$Ni$_{x}$BO$_{5}$\cite{bezmaternykh2014} or Pr$_{3}$Fe$_{3}$Sb$_{7}$~\cite{Pabst2023}. This can be described by a model which considers two antiferromagnetically interacting subsystems, each being ferrimagnetically ordered. At a slightly larger field of 0.01 T we observe the same behavior but the field-polarized component is larger. Small kinks and weak hysteresis are observed at fields up to 0.35 T, as shown in the same figure.

\begin{figure}
\begin{center}
\includegraphics[width=0.8\columnwidth]{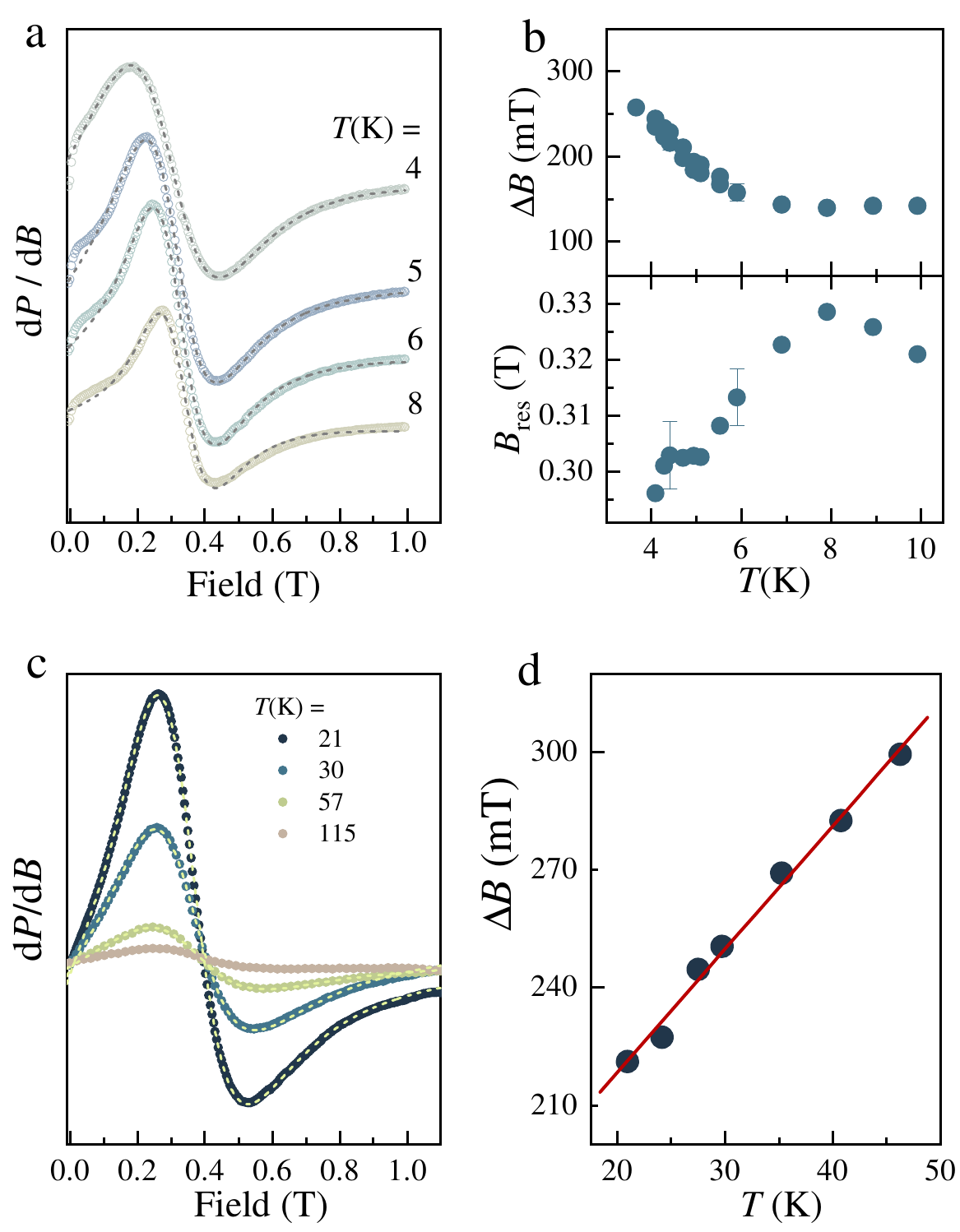}
\end{center}
\caption{ESR spectra $\mathrm{d}P/\mathrm{d}B$ (symbols) of Eu$_{10}$Hg$_{55}$ at different temperatures and Lorentzian line fittings (dashed lines) resulting in ESR linewidth, $\Delta B$, and resonance field $B_{\rm res}$. ESR results shown in (a) and (b) refer to single crystalline, (c) and (d) to powdered Eu$_{10}$Hg$_{55}$ from the same batch. Solid line shown in (d) represents the best fit to Eq.(\ref{korringa}).}
\label{FigESR1}
\end{figure}

Electron Spin Resonance (ESR) was employed as a local probe technique to characterize the magnetic properties of Eu in Eu$_{10}$Hg$_{55}$. In general, an Eu$^{2+}$ ion in the magnetic $4f^{7}$ configuration ($L = 0, S = J = 7/2$) can be easily detected by ESR, whereas an Eu$^{3+}$ ion with a non-magnetic ($J = 0$) $4f^{6}$ configuration is expected to be ESR-silent.\cite{sichelschmidt05b, kokal14a, aydemir16a} Typical ESR spectra of Eu$_{10}$Hg$_{55}$ are shown in Figs.~\ref{FigESR1} (a) and (c) for selected temperatures. We have investigated the ESR of a single crystal as well as of a powder sample, both originating from the same batch (Sample 3), in two separate temperature regions. It turned out that single crystal ESR data are meaningful only at low temperatures, $T < 10$ K, where the microwave penetration depth is large enough, allowing the intensity of the sample signal to exceed that of the background signal. The anisotropy upon rotating the crystal in the external magnetic field was very weak, showing no resolvable changes in the linewidth and resonance field. Above 20 K the ESR examination of powder ensures that, despite high electrical conductivity (small microwave penetration depth), the most of the sample volume contributes to the ESR, resulting in a stronger ESR signal. Then, the ESR line was visible up to $\simeq50$ K, above which the line broadening combined with the decrease of the spectra amplitude made it difficult to reliably fit the ESR spectra. 

The results of the single crystal ESR investigation are compiled in Figs.~\ref{FigESR1} (a) and (b), those of the powder in Figs.~\ref{FigESR1} (c) and (d). The best fit of the spectra to a single, asymmetric Lorentzian ("Dysonian") line (see Supporting Information) is indicated by the dashed lines revealing an asymmetry that is increasing with decreasing temperature due to a reduced skin depth or increased sample conductivity towards low temperatures. This fitting yields the parameters, which are given in Figs.~\ref{FigESR1} (b) and (d). As shown in Fig.~\ref{FigESR1} (b), in the low temperature region the linewidth $\Delta B$ shows a continuous broadening below $T\simeq6$ K, indicating the enhancement of Eu$^{2+}$ spin correlations when approaching magnetically ordered phases below 5 K. At the same time, the resonance field $B_{\rm res}$ shows a pronounced temperature dependence. This points to the build-up of internal fields, as expected in the vicinity of magnetic order. 

For $T > 10$ K, the resonance field $B_{\rm res}$ has a weak temperature dependence and is slightly increasing with increasing temperature. The obtained value of the $g$ factor at 40 K, $g = 2.01$, is close to the value $g_{0} = 1.9935$, expected for an Eu$^{2+}$ ion in a crystalline field environment of cubic symmetry.\cite{abragam70a} This suggests that the observed ESR spectra are due to localized magnetic moments of Eu$^{2+}$ ions in Eu$_{10}$Hg$_{55}$. The ESR intensity is proportional to the magnetic susceptibility of the ions, which produce the ESR signal. In Eu$_{10}$Hg$_{55}$, the dominant contribution to magnetic susceptibility is due to Eu$^{2+}$ ions (see discussion above). The ESR intensity shows a Curie-Weiss-like behavior, qualitatively similar to the bulk magnetic susceptibility data. This confirms that the observed ESR signal in Eu$_{10}$Hg$_{55}$ comes from localized magnetic moments of Eu$^{2+}$ ions. The ESR linewidth $\Delta B$ provides information on the local spin dynamics of the resonant magnetic moments. In this respect important information is obtained from the temperature dependence of the linewidth plotted in  Fig.~\ref{FigESR1}(d), showing a linear thermal broadening. This indicates the dominant role of a Korringa relaxation of the localized Eu$^{2+}$ moments via scattering off the conduction electrons:

\begin{equation}
\Delta B(T) = \frac{\pi k_{B}}{g \mu_{B}}(J_{fce}N(E_{F}))^{2}T = bT
\label{korringa}
\end{equation}

\noindent where $J_{fce}$ is the exchange constant between the Eu$^{2+}$ 4$f$ localized magnetic moments and the conduction electrons, $N(E_{F})$ is the conduction electron density of states at Fermi Energy, and $b$ is the Korringa slope.\cite{barnes81a} The obtained value in Eu$_{10}$Hg$_{55}$ is $b = 3$ mT/K, which is significantly larger than the typical value of $b \approx 1$ mT/K of the $S$ state 4$f^{7}$ local moments in conventional metals.\cite{barnes81a} According to Eq.(\ref{korringa}), the large Korringa slope in Eu$_{10}$Hg$_{55}$ indicates a strong coupling of the Eu$^{2+}$ localized magnetic moments with conduction electrons and/or a large density of state at the Fermi level.

It is well known that applied pressure tends to suppress the Eu$^{2+}$ state in favor of the smaller-volume Eu$^{3+}$ state.\cite{Yannello2019, Tan2016} To explore such a possibility, we probed magnetic properties of Eu$_{10}$Hg$_{55}$ under applied pressure in a diamond anvil cell (DAC). A few small crystals of the material were loaded into the DAC, with the culet diameter of 0.5 mm. Small amounts of ruby and Nujol oil were added as pressure indicator and pressure-transmitting medium, respectively. The pressure was applied as force measured in kN, while the actual pressure was determined by measuring a shift in the ruby fluorescence peak: $P(\text{GPa}) = 175.2[(\lambda/\lambda_0)^{10.7} - 1]$, \cite{Chijioke2005} where $\lambda$ and $\lambda_0$ are wavelengths of maximum fluorescence observed under applied and ambient pressure, respectively. The magnetization data were corrected by subtracting the background measured on an empty DAC. The field-dependent magnetization measured at 10 K under ambient pressure and at 12.8 GPa did not reached saturation and did not show any anomalies, as expected for paramagnetic behavior above the ordering temperature. A comparison of the magnetization data normalized against the ambient-pressure curve (Fig.~\ref{fig:PhaseD}(c)) revealed $\sim 5$\% decrease in the maximum magnetization at 7 T, suggesting that the applied pressure results in partial suppression of the Eu$^{2+}$ state and increase in the average oxidation state of Eu. More detailed studies are planned in the future, combining magnetic measurements and X-ray Absorption Near Edge Structure (XANES) spectroscopy to understand the pressure-dependent electronic and magnetic behavior of this material.

Ambient-pressure XANES at the Eu L$_{3}$-edge was used to probe the oxidation state of Eu. The intense peak at the onset of the absorption, referred to as the white line, is found at 6975 eV for Eu$^{2+}$ and at 6983 eV for Eu$^{3+}$, making the two oxidation states easily distinguishable \cite{Gschneidner2004}. The use of the High-Energy-Resolution Fluorescence-Detected (HERFD) mode to collect XANES results in sharper spectral features and increases considerably the sensitivity to the oxidation states of Eu.\cite{Glatzel2013} The HERFD XANES at the Eu L$_{3}$-edge was collected on sample 1 at the ROBL \cite{ROBL, Kvashnina2016} beamline at the ESRF (Fig.~\ref{fig:PhaseD}(b)). The sample was manipulated and mounted in the sample holder in an Ar glove-box. However, the kapton window in front of the sample-holder used to allow x-rays to enter and exit was not sufficient to prevent air to penetrate. In the HERFD XANES spectrum, shown in Fig.~\ref{fig:PhaseD}(b), the peak corresponding to Eu$^{3+}$ is dominating over the one of Eu$^{2+}$. The peaks were fitted with two Split-Voigt functions to account for the asymmetry. Quantification based on the ratio of the areas below the fitted peaks indicate 0.2 Eu$^{2+}$ and 0.8 Eu$^{3+}$.   

\section*{Conclusions}

In this work, we present the analysis of complex magnetic order of Eu$_{10}$Hg$_{55}$, which has four distinct crystallographic positions for Eu atoms. It appears that in this material, both Eu$^{2+}$ and Eu$^{3+}$ speciesmight be present, which is made possible by the cage-like structure of this compound -- polyhedra of slightly different volumes are likely hosting different species. The magnetic order of Eu$_{10}$Hg$_{55}$ is fairly fragile -- probably also a result of the structure, in which Eu moments are diluted. The exact magnetic structure assignment is hampered by the fact that both neutron diffraction studies and theoretical analysis are likely impossible for Eu$_{10}$Hg$_{55}$. Given that the isostructural Ca$_{10}$Hg$_{55}$ and Sr$_{10}$Hg$_{55}$ compounds exist and show superconductivity \cite{Nixon2025}, partial substitution of Eu by either Ca or Sr is of interest and is currently being pursued. If an analysis of spin textures in Eu$_{10}$Hg$_{55}$ can be carried out in an air-free atmosphere, this would also be of interest.

\section*{Acknowledgments}

Authors acknowledge useful discussions with C. Hoch and Yu. Grin. ES is grateful for the support of the Christiane N\"usslein-Volhard-Stiftung. ES, NZ and MK acknowledge the support of the Boehringer Ingelheim Plus 3 Program. NZ is grateful for the support of the Humboldt Foundation through the Philipp Schwartz Initiative. RN and ES acknowledge the funding from the International Max Planck Research School for Chemistry and Physics of Quantum Materials and the funding provided by the German Research Foundation (DFG-528628333 “Complex compounds based on mercury”). AS acknowledges the funding by the Shota Rustaveli National Science Foundation of Georgia (SRNSFG) under grant no. FR-23-560. MS and SAA acknowledge support by the U.S. National Science Foundation (DMR-2233902) and the NSF-MRI program for supporting acquisition of the Quantum Design MPMS-3 system used for magnetic measurements (DMR-2216125). Cover design: Christina Pouss, MPI CPfS.

\section*{Conflict of Interest}

Please enter any conflict of interest to declare.

\setlength{\bibsep}{0.0cm}

\bibliography{EuHg_references.bib}

\clearpage

\section*{Supporting Information}

\renewcommand{\thetable}{S\arabic{table}}
\renewcommand{\thefigure}{S\arabic{figure}}
\setcounter{figure}{0}
\setcounter{table}{0}

\section{Synthesis}

Many issues arise from experimental work with mercury, including high vapor pressure, high chemical reactivity, toxicity, and extreme air sensitivity. To mitigate these effects, a specialized laboratory environment is needed \cite{LeitheJasper2006}. As we have previously shown \cite{Witthaut2023, Prots2022U, Prots2022La, Svanidze2019}, it is possible to obtain previously inaccessible intrinsic crystallographic and physical property data on mercurides and amalgams.

All samples were synthesized by combining Hg (droplet, Alfa Aesar, 99.999\%) with Eu (pieces, Alfa Aesar, 99.9\%) with a ratio of Eu:Hg of 5:95. Samples were sealed in tantalum tubes to preserve stoichiometry. To protect samples from air and moisture, all syntheses were performed in an argon-filled glove box system. The heating profiles consisted of heating to 500 $^\circ$C and then slowly cooling down to room temperature. The samples were then placed in specialized crucibles \cite{Crucibles} and centrifuged at room temperature to remove the excess mercury. This method can eliminate most mercury, any remainder can be removed by allowing the crystals to sit on gold foil for several days.

\begin{figure}[ht!]
\includegraphics[width=\columnwidth]{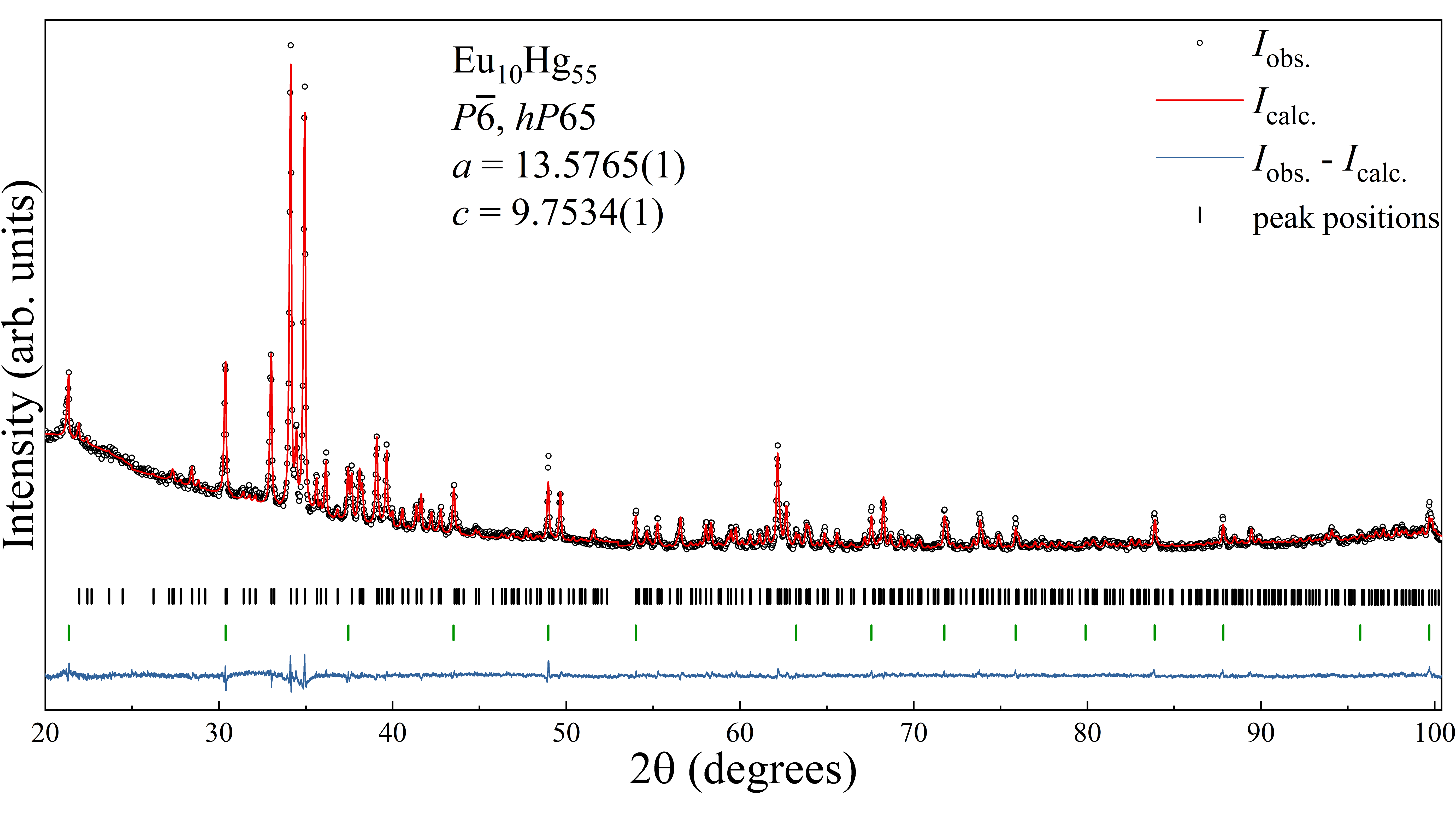}
\caption{Powder XRD pattern for Eu$_{10}$Hg$_{55}$ with LaB$_6$ as an internal standard (I$_{obs.}$, symbols), together with the calculated profile (I$_{calc.}$, red line), difference between them (I$_{obs. - Icalc.}$, blue line), and calculated positions of the Bragg reflections (vertical ticks). Eu$_{10}$Hg$_{55}$ phase is marked as black ticks, whereas LaB$_6$ is marked as green.} 
\label{fig:Eu10Hg55_xrd}
\end{figure}

Powder X-ray diffraction was performed on a Huber G670 Image plate Guinier camera with a Ge-monochromator (CuK$\alpha_1$ radiation, $\lambda$ = 1.54056 \AA) using LaB$_6$ as an internal standard. Powders were sealed between two Kapton films to prevent oxidation and decomposition. Phase identification was done using the Match 3! software\cite{Putz2023}. The WinCSD software package\cite{Akselrud2014} was used for lattice parameters determination by a least-squares refinement using the peak positions, extracted by profile fitting, and for crystallographic analysis. Single crystal diffraction was not carried out due to the extreme air-sensitivity of these materials. The values of the lattice parameters, refined from powder X-ray data (see Fig.\ref{fig:Eu10Hg55_xrd}), were in agreement with those reported previously\cite{Tambornino2015}, as summarized in Fig.~\ref{fig:Structure}.

\section{Electron spin resonance (ESR) analysis}

In our ESR measurements we used a single crystal (arbitrarily oriented) and powder of Eu$_{10}$Hg$_{55}$ from the same batch (NZs-020-1). Both samples were mounted under Ar gas atmosphere in quartz sample tubes and fixed by paraffin.
We used a continuous-wave ESR setup at X-band frequency ($\nu = 9.4$ GHz) equipped with a continuous He gas-flow cryostat in the temperature range $3 < T < 300$ K.
The resonance line was recorded as the field derivative of the absorbed microwave power $P$ to enable an improvement of the signal to noise ratio by a lock-in phase sensitive detection. 
The resonance field ($B_{\rm res}$, determined by the Eu $g$ value and internal magnetic fields), linewidth ($\Delta B$, determined by the Eu spin dynamics) and intensity (being a measure of the local susceptibility of the Eu spins) of the ESR signals were obtained from fitting the $\mathrm{d}P/\mathrm{d}B$ spectra with the first derivative of a Lorentzian shape: \cite{joshi04a,rauch15a}

\begin{equation}
\frac{\mathrm{d} P}{\mathrm{d}B} \sim \frac{\mathrm{d}}{\mathrm{d}B}\left[ \frac{\Delta B + \alpha(B-B_{\rm res})}{\Delta B^{2} + 4(B-B_{\rm res})^{2}}+\frac{\Delta B - \alpha(B+B_{\rm res})}{\Delta B^{2} + 4(B+B_{\rm res})^{2}}\right]
\label{lineshape}
\end{equation}

This lineshape, often referred to as Dysonian, contains a parameter $\alpha$ ($0 \leq \alpha \leq 1$) indicating the mixture of absorptive and dispersive components of the susceptibility due to the finite depth of microwave penetration in a conducting sample. If the penetration depth is small in comparison to the sample grain size, $\alpha$ approaches 1. 

\section{Analysis of magnetic properties}

Magnetic properties were examined using a Quantum Design (QD) Magnetic Property Measurement System in the temperature range of 1.8–300 K and under various applied magnetic fields. Samples were sealed in glass tubes to prevent reaction with air. The specific heat data were collected on a QD Physical Property Measurement System from 0.4 K to 10 K and under various applied magnetic fields. Due to the high air-sensitivity of the Eu$_{10}$Hg$_{55}$ samples, they were covered with Apiezon N vacuum grease, the background contribution of which was subtracted. It was not, however, possible to measure electrical resistivity of Eu$_{10}$Hg$_{55}$, due to its high air-sensitivity, coupled with tendency to form mercury thin films which then dominate the transport behavior. The morphology of Eu$_{10}$Hg$_{55}$ crystals (see Fig.~\ref{fig:Structure}) did not allow for the implementation of the previous experimental solution for measurement of electrical resistivity, developed by our group \cite{Witthaut2023, Prots2022La, Prots2022U}.

\section{X-ray absorption measurements}

HERFD XANES \cite{Glatzel2013} spectra at the Eu L$_{3}$-edge were acquired at the ROBL beamline of the ESRF \cite{ROBL}. The incoming beam was monochromatized with a fixed-exit Si(111) double crystal monochromator. The Eu La1 characteristic fluorescence was analyzed with an X-ray emission spectrometer \cite{Kvashnina2016} based on Rowland geometry on which 4 Ge(333) crystal analyzers were mounted. The HERFD XANES were obtained by collecting the maximum of the Eu La1 emission line with a bandwidth of $\sim$1.5 eV.

Sample 1 was mounted on a sample holder in an Ar glovebox. The sample holder was sealed with a kapton tape to allow X-rays to enter and exit. The sample was sealed in a bottle and transported to the beamline. The bottle was open and the sample placed on the sample stage. Considering the time required to interlock the hutch and briefly scan the sample for centering, we estimate that the first HERFD XANES was acquired after 3 to 5 minutes after the sample holder was put in air. This time was unfortunately sufficient to oxidize a large amount of the Eu$^{2+}$. 

\end{document}